\documentstyle[epsf]{l-aa}
\newcommand{\doppler}{\delta}
\newcommand{\bmag}{b}
\newcommand{\tcross}{t_{\rm cr}}
\newcommand{\tvar}{t_{\rm var}}
\newcommand{\tvarfive}{t_5}
\newcommand{\nubreak}{\nu_{{\rm b,}15.3}}
\newcommand{\dist}{D}
\newcommand{\flux}{F}
\newcommand{\nusync}{\nu_{{\rm s,}18}}
\newcommand{\nucompt}{\nu_{{\rm c,}27}}
\newcommand{\tesc}{t_{\rm esc}}
\newcommand{\sigmaT}{\sigma_{\rm T}}
\newcommand{\melec}{m_{\rm e}}
\newcommand{\nelec}{n_{\rm e}}
\newcommand{\ngamma}{n_{\gamma}}
\newcommand{\db}{\ell_{\rm B}}
\newcommand{\elcomp}{\ell_{\rm e}}
\newcommand{\lowq}{q_{\rm e}}

\newcommand{\gammamax}{\gamma_{\rm max}}
\newcommand{\gammabr}{\gamma_{\rm br}}
\newcommand{\elinject}{Q_{\rm e}}

\newcommand{\dopplermn}{\delta_{15}}

\newcommand{\eqb}{\begin{eqnarray}}
\newcommand{\eqe}{\end{eqnarray}}

\begin{document}
\thesaurus{11.01.2; 11.02.2 Mkn~421; 11.10.1; 13.07.3; 13.25.3}
\title{Variability in the synchrotron self-Compton model  
of blazar emission}
\author{A. Mastichiadis, J.G. Kirk}
\institute{Max-Planck-Institut f\"ur Kernphysik,
Postfach 10 39 80, D-69029 Heidelberg,
Germany}
\offprints{J.G. Kirk}
\date{Received \dots Accepted \dots}
\maketitle
\begin{abstract}
We present a model of the spectra of gamma-ray emitting blazars 
in which a single homogeneous emission region both emits synchrotron
photons directly and scatters them to high (gamma-ray) energy before
emission (a \lq synchrotron self-Compton\rq\ or SSC model).
In contrast to previous work, we follow the full time dependent 
evolution of the electron and photon spectra, assuming a power-law
form of the electron injection and examine the predictions
of the model with regard to variability of the source.
We apply these computations to the 
object Mkn~421, which displayed rapid variability in its 
X-ray and TeV emission during a multiwavelength 
campaign in 1994.
This observation strongly implies that the same population of electrons 
produces the radiation in both energy bands. 
By fitting first the observed quiescent spectrum
over all 18 orders of magnitude in frequency, 
we show that the time dependence of the 
keV/TeV flare could have been the result 
of a sudden increase in the maximum 
energy of the injected electrons. 
We show also that different types of flare may occur
in this object and others, and 
that the energy band most sensitive to the properties of 
the acceleration mechanism is the X-ray band.
\end{abstract}

\section {Introduction}

Mkn~421 is a nearby (z=0.03) BL~Lac object, with emission characterised 
by variability at all wavelengths. It appears to emit purely 
non-thermal continuum radiation; emission lines are absent. 
In addition, Mkn~421 
is the first extragalactic source which has confirmed detections in TeV 
$\gamma-$rays (Punch et al. \cite{punchetal92}, Schubnell et al. \cite{schubnelletal96}).
It is also one of the $\sim$50 active galactic nuclei 
(AGN) to have been detected by the EGRET experiment on the Compton 
Gamma-Ray Observatory (von Montigny et al.~\cite{vonmontignyetal95}).

Concerted efforts have been undertaken to  obtain simultaneous 
observations of blazars over a wide spectral range. 
In April and 
May~1994, Mkn~421 was the target of such an observation (Macomb et 
al.~\cite{macomb95}, \cite{macomb96}). This object had been observed in previous 
multi-wavelength campaigns (Brodie, Bowyer \& Tennant~\cite{brodieetal87},
Makino et al.~\cite{makinoetal87})
but the 1994 campaign was the first to cover 
both the GeV and TeV gamma-rays as well as the radio to X-ray bands. 
Furthermore these coordinated observations were fortunate to coincide 
with an increase in the flux above 250 GeV by a factor of several in a 
few days. Contemporaneous observations performed by ASCA showed the 
X-ray flux was also in a high state compared to earlier observations. 
Significantly, EGRET did not detect a flare in the MeV-GeV regime, nor 
was any change observed in the radio to optical bands. These results 
imply that a single population of electrons is responsible for the 
radiation in both the keV and TeV bands and provide a potential test for 
the theoretical models that have been put forward to explain the 
emission from AGN (e.g., 
Maraschi, Ghisellini \& Celotti~\cite{maraschietal92}, 
Dermer \& Schlickeiser~\cite{dermerschlickeiser93}, 
Sikora, Begelman \& Rees~\cite{sikoraetal94}, 
Blandford \& Levinson~\cite{blandfordlevinson95},
Marcowith, Henri \& Pelletier~\cite{marcowithetal95}).                

One of the most promising models for the emission mechanism of BL~Lac 
objects (including Mkn~421) is the \lq\lq synchrotron self-Compton 
model\rq\rq\ (SSC). 
Originally applied to nonthermal emission of 
frequency up to X-rays (Jones, O'Dell \& Stein \cite{jonesetal74}),
this mechanism has been revised to 
accommodate the recent discoveries by EGRET and WHIPPLE. The generic 
picture is that the radio to X-ray photons arise as the synchrotron 
radiation of relativistic electrons accelerated by some unspecified 
mechanism in a blob of plasma which itself moves relativistically 
outwards from the core of the AGN; the gamma-rays are the result of the 
inverse Compton scattering of the synchrotron photons by the 
relativistic electrons in the blob. Several expositions of the steady 
state spectrum predicted by this model can be found in the literature. 
Variability is also predicted in these models, where it arises
not from a non-steady electron distribution but from changes
in the parameters of the source as it moves through the jet
(Celotti, Maraschi \& Treves~\cite{celottietal91},
Maraschi, Ghisellini \& Celotti~\cite{maraschietal92}, Bloom \&
Marscher~\cite{bloommarscher96}).  

However, during a flare, the electrons may not have time to reach a steady 
state since their cooling time can be much longer than the timescale of 
the flare. This in turn means that in order to follow the evolution of a 
multifrequency flare such as detected during the 1994 campaign, time 
dependent calculations, taking into account electron injection and 
cooling and following the evolution of  the spectrum, are indispensable. 
In this paper we present such a calculation. Section~2 contains a brief 
description of the computational method; in Sect.~3 we compare the predictions
of our model to the 
multiwavelength spectrum of Mkn~421 and show the time dependent 
evolution of flares obtained by varying certain parameters in the 
injected electron spectrum. We conclude in Sect.~4 with a discussion of our 
results. 

\section{Computational model}
As part of a general approach to the modelling of AGN spectra, including 
the self-consistent acceleration of a hadronic component, we have 
developed a numerical code capable of following the time evolution not 
only of the proton, but also of the photon and electron distributions in 
the source (Mastichiadis \& Kirk \cite{mastichiadiskirk95}, henceforth MK95). 
To accomplish this, we assumed 
each component can be represented by a spatially averaged distribution, 
treating escape from the source by a simple \lq catastrophic 
loss\rq\ term. The system of three time-dependent kinetic equations for 
the distributions as functions of energy is then amenable to numerical 
solution. The relevant physical processes taken into account for the 
electron and photon distributions include synchrotron radiation, inverse 
Compton scattering both in the Thomson and Klein Nishina regimes, as 
well as photon-photon pair production and synchrotron self-absorption. 
Other processes included in the original code, such as pair 
annihilation and photon downscattering on cold electrons, turn out 
to be irrelevant for the case we are considering here. 

Two minor modifications must be made before this code can describe the 
SSC model for blazars. Firstly, since the acceleration mechanism is not 
directly addressed, it is not necessary to follow the proton 
distribution. Instead, an arbitrary dependence of the electron injection 
function on energy and time can be implemented. Secondly, the extremely 
high flux of gamma-rays observed from blazars indicates that the source 
must be Doppler-boosted. Thus, quantities such as the photon spectrum, 
which are computed in the rest frame of the source, require a Lorentz 
transformation into the observer's frame. 

With these changes, the parameters which specify the source are as 
follows: 
\begin{itemize} 
\item 
the Doppler factor $\doppler=1/[\Gamma(1-
\beta\cos\theta)]$, where $\Gamma$ and $c\beta$ are the Lorentz factor 
and speed of the blob, and $\theta$ is the angle between its direction 
of motion and the line of sight to the observer, 
\item 
the radius $R$ of 
the source (in its rest frame, in which it is assumed spherical) 
from which the crossing 
time $\tcross=R/c$ can be defined; the variation timescale 
in the galaxy frame is then given by $\tvar=R/(\doppler c)$, 
\item 
the magnetic 
field strength $B$, specified in units of the critical magnetic
field: $\bmag=B/(4.414\times10^{13}\,{\rm G})$.
When combined with $R$, this determines the magnetic \lq\lq compactness\rq\rq\ 
which can be defined in analogy with the photon compactnesss (MK95) as
$\db=\sigmaT R B^2/(8\pi \melec c^2)$, 
where $\sigmaT$ is the Thomson cross section and $\melec$ the 
electron rest mass, 
\item 
the electron injection spectrum, for which we take 
$\elinject=\lowq \gamma^{-s}e^{-\gamma/\gammamax}$ where $\gamma$ is the 
electron Lorentz factor. The three scalar parameters used to specify 
this distribution in the following are $s$, $\gammamax$ and the electron
injection compactness $\elcomp={{1\over 3}\melec c\sigmaT R^2} 
\int_1^\infty {\rm d}\gamma (\gamma-1) Q_{\rm e}$ (see MK95)
\item 
the effective escape time $\tesc$ 
of relativistic electrons, which can be identified as the timescale over 
which adiabatic expansion losses limit the accumulation of relativistic 
electrons within the source. 
\end{itemize} 

In terms of these parameters, the equation governing the evolution of 
the electron distribution $\nelec$ is:
\eqb
{\partial \nelec(\gamma,t)\over \partial t}
+{\nelec(\gamma,t)\over\tesc}&=&
Q_{\rm e}(\nelec,\ngamma,\gamma,t)+L_{\rm e}(\nelec,\ngamma,\gamma,t)
\label{ekinetic}
\eqe
Here $L_{\rm e}$ denotes the electron loss terms (i.e., synchrotron 
losses and inverse Compton scattering), while $Q_{\rm e}$ 
is the 
injection term. In all cases discussed in this paper the contribution 
of photon--photon pair production to both photon absorption and 
electron injection is negligible, so that $Q_{\rm e}$ is essentially a 
function of only $\gamma$ and $t$.
The normalisation used is one in which time is 
measured in units of the light crossing time of the source (of size $R$) 
and the particle density refers to the number contained in a volume element of 
size $\sigmaT R$. All
quantities are measured in the rest frame of the blob.
The corresponding equation for the photon distribution $\ngamma$ reads:
\eqb
{\partial \ngamma(x,t)\over \partial t}
+\ngamma(x,t)
&=&Q_{\gamma}(\ngamma,\nelec,x,t)+L_{\gamma}(\ngamma,\nelec,\gamma,t)
\label{gkinetic}
\eqe
where here $Q_{\gamma}(\ngamma,\nelec,x,t)$ 
represents the source terms for  photons 
of frequency $\nu=x \melec c^2/h$ (i.e., synchrotron radiation and 
inverse Compton  scattering) and
$L_{\gamma}(\ngamma,\nelec,\gamma,t)$ the loss term, which arises from 
photon-photon pair production and is 
negligibly small in the current application. 
The source and loss terms are discussed 
in detail in MK95.
Using the above parameters in the code it is possible
to fit to the multiwavelength spectrum as reported
in Macomb et al.~(\cite{macomb95}, \cite{macomb96}) during both the quiescent and the
flaring stages.

\begin{figure}[t]
\epsfxsize=10.2 cm
\epsffile{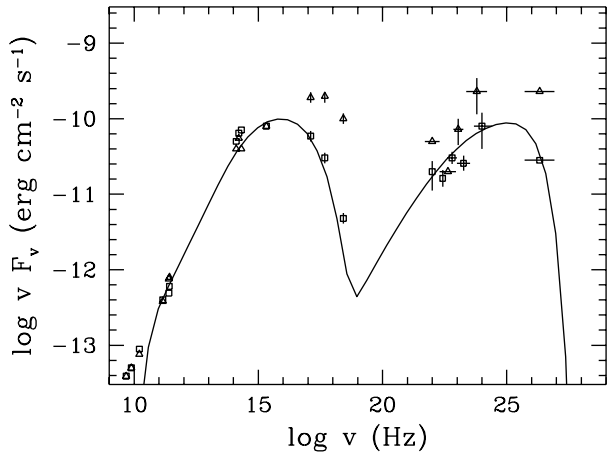}
\vspace{-2.5 cm}
\caption{\protect\label{quiet}
The predicted and observed emission from Mkn~421 during its quiescent state.
Squares indicate observations during the quiescent phase. Triangles are
observations during the flaring state 
(Macomb et al.~\protect\cite{macomb96}).
For the adopted parameters see text.}
\end{figure}

\section{Fits to the multiwavelength spectrum of Mkn 421}
\subsection{The quiescent state}
The seven parameters $\elcomp$, $s$, $\gammamax$, $\doppler$, $R$, $b$, 
and $\tesc$ are strongly constrained by the observations. In the case of 
Mkn~421, both the X-ray emission and the TeV emission are rapidly 
variable (see Macomb et al.~\cite{macomb95},
Takahashi et al.~\cite{takahashietal96}, 
Schubnell et al. \cite{schubnelletal96}). 
Radio frequency photons, however, exhibit a quiescent                   
spectrum. Since the time required for electrons to cool by synchrotron 
emission to energies at which they emit substantial numbers of 
infra-red photons is excessively long, we 
assume that either adiabatic expansion 
of the blob or escape from the source intervenes to limit the time over 
which a given electron radiates in this frequency range. Then the 
observed infra-red to radio spectral index $\alpha\approx 0.35$ is directly 
related to $s$ by the well-known formula $\alpha=(s-1)/2$. 

Although variable, the general form of both the X-ray and gamma-ray 
spectra indicates that in the SSC model synchrotron photons are emitted 
up to a maximum frequency of approximately $10^{18}\,$Hz while inverse Compton 
photons are present up to at least $10^{27}\,$Hz. Denoting these 
frequencies by $\nusync\times10^{18}\,$Hz and 
$\nucompt\times10^{27}\,$Hz, respectively, we have 
\eqb 
\doppler\bmag\gammamax^2\approx 10^{-3}\nusync
\label{sbreak}
\\
\doppler\gammamax\approx 3\times 10^{6} \nucompt
\label{cbreak}
\eqe
(MK95).
Eq.~(\ref{cbreak}) assumes that photons of frequency $\nucompt$ 
are produced by inverse Compton scatterings in the Klein-Nishina regime.
   From these expressions
we can immediately deduce $\gammamax$ and 
the magnetic field strength in terms of the (as yet undetermined)
Doppler factor $\doppler$: 
\eqb
\gammamax&=& 3\times 10^6 \nucompt \doppler^{-1}
\\
B&=& 5\times 10^{-3} \doppler\nusync\nucompt^{-2} \ {\rm gauss}
\label{bequation}
\enspace.
\eqe

The next step is to relate the observed
bolometric nonthermal flux $\flux$ of the 
source to the power injected in relativistic electrons. 
Using the normalisation of MK95
we require the injected electron compactness $\elcomp$
in the rest frame of the source to be
\eqb
\elcomp
&=&
{{3\sigmaT\flux\dist_{\rm L}^2}\over{\doppler^4 R \melec c^3}}
\label{elcomp}
\eqe
where $\dist_{\rm L}$ is the luminosity distance to the source 
given by $\dist_{\rm L}=2c[z+1-(z+1)^{1/2}]/H_0$ for a $q_0=1/2$,
$\Lambda=0$ cosmology; for the Hubble constant we assumed
$H_0=75~h~{\rm{km}~s^{-1}~Mpc^{-1}}$ and the redshift of Mkn~421 
is $z=0.0308$.
Observations of Mkn~421 indicate that in the SSC model, the flux 
of synchrotron photons (i.e., those of frequency $<\nusync$) is 
comparable to the flux of inverse Compton photons (those with 
frequency $>\nusync$).
In the rest frame of the blob, this implies
approximate equality of the two compactnesses $\elcomp$ and $\db$ since
these quantities are proportional to the internal photon and magnetic
energy  density respectively.
Writing $\eta=\elcomp/\db$ we have
\eqb
R&=&3 \times 10^{19} \elcomp\eta^{-1} B^{-2}\ {\rm cm}
\label{rle}
\eqe
with $B$ in gauss. Using Eq.~(\ref{elcomp}) and the observationally derived
relation $F\dist^2_{\rm L}\approx 6\times 10^{44}$ erg/sec, 
we obtain from Eq.~(\ref{rle})
\eqb 
R\simeq1.2\times 10^{18}\left(
{{F\dist^2_{\rm L}}\over {6\times 10^{44}~\rm 
{erg/sec}}}\right)^{1/2}
\eta^{-1/2}B^{-1}\doppler^{-2}~\rm {cm}.
\label{radius}
\eqe

The observation that
the synchrotron spectrum steepens at a frequency 
somewhere between millimeter and optical wavelengths 
(Brodie, Bowyer \& Tennant~\cite {brodieetal87})
enables one to estimate the escape time
first by calculating an effective Lorentz factor $\gammabr$ below
which escape is faster than cooling
\eqb
\gammabr\simeq {3\over{4\db\tesc}}
\eqe
and then relating this to the turnover frequency by 
\eqb
\nu_{\rm b}\simeq 1.3\times 10^{20}\doppler b\gammabr^2.
\eqe
This approach gives an escape time
\eqb
\tesc&=8.9\times 10^{14}\doppler^{1/2}B^{-3/2}R^{-1}\nubreak^{-1/2}
\label{break}
\eqe
expressed in units of $\tcross$. Here $\nubreak=\nu_{\rm 
b}/10^{15.3}\,{\rm Hz}$ is the 
frequency of the spectral break.

Finally, the rapid variability of the 
X-ray and TeV emission
on timescales $\tvar$ of $10^5\,$secs constrains the overall size of the 
source and the Doppler boosting factor. Thus using $\tvar=R/(c\doppler)$
and eliminating $B$ between Eqs.~(\ref{bequation}) and (\ref{radius})
to find the value of $R$, we deduce
\eqb
\doppler= 15\tvarfive^{-1/4}
\label{dopplimit}
\eqe
where $\tvarfive=\tvar/(10^5\,{\rm s})$ and we have inserted canonical 
values for the observed quantities. 

Following the above guidelines one can 
allow the code to reach a stationary state and
find detailed fits to the 
\lq quiescent\rq\ multiwavelength spectrum of Mkn~421 (Macomb et 
al.~\cite{macomb95}, \cite{macomb96}). With the exception of the slope of injected 
electrons $s$, all other parameters can be expressed in terms of the 
Doppler factor $\doppler$. Thus, we first use the 
approximate observed quantities $\nusync$, $\nucompt$, $F\dist^2_{\rm 
L}$, $\alpha$, $\nubreak$, and the observed approximate
equality of $\elcomp$ and $\db$, to fit the observed
quiescent spectrum and subsequently adjust $\delta$ to fit the observed 
variation time $\tvarfive$. 
We find a satisfactory fit using 
the parameters $R=4.7~10^{16}\dopplermn^{-3}$ cm, 
$B=.07\dopplermn$ gauss, $\gammamax=2~10^5\dopplermn^{-1}$, $s=1.7$, 
$\elcomp=1.93~10^{-5}\dopplermn^{-1}$ and 
$\tesc=3.3\tcross\dopplermn^2$. Here $\dopplermn=\doppler/15$. 
Fig.~\ref{quiet} shows the calculated quiescent spectrum for
$\dopplermn=1$.

\begin{figure}[t]
\epsfxsize=10.2 cm
\epsffile{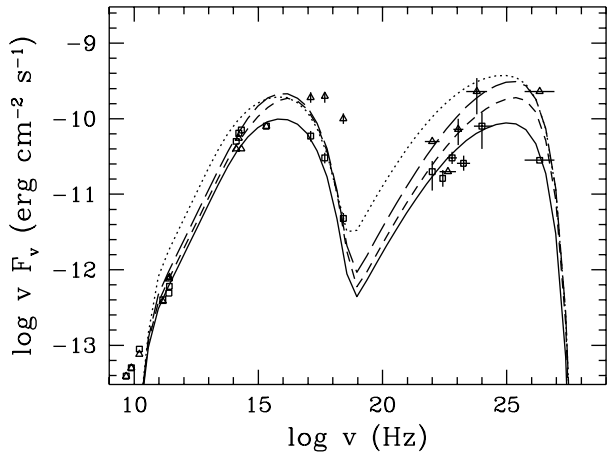}
\vspace{-2.5 cm}
\caption{\protect\label{qinjplot}
Evolution of the multiwavelength spectrum of Mkn 421 in the case where the 
amplitude of the electron injection changes impulsively by a factor of 3.
The solid line corresponds to the quiescent spectrum of Fig.~1. The short
and long dashed
lines show the spectrum at 1 and $2\protect\tvar$ after the change in the electron
injection (corresponding to 1.2 and 2.4 days respectively).
The dotted line shows the new steady state which, however, is
achieved many \protect$\tvar$ later.}
\end{figure}
\begin{figure}
\epsfxsize=10. cm
\epsffile{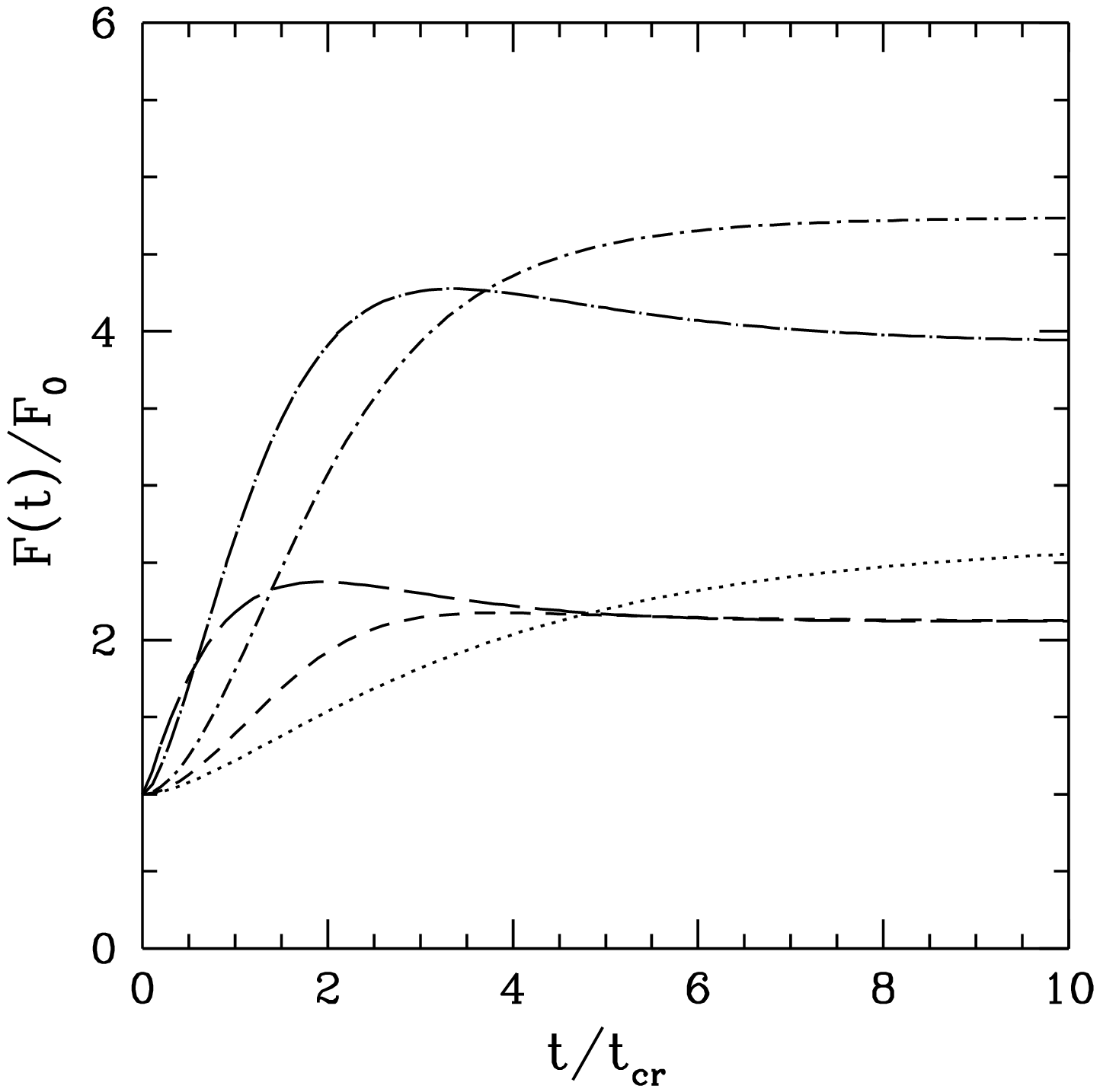}
\caption{\protect\label{lightqinj}
Plot of the flux at various frequencies (normalised to its quiescent value)
for the flare that corresponds to a change in \protect$\lowq$ by a factor of 3
over the quiescent state. The dotted line corresponds to a  wavelength of 0.4 cm,
the small dash line to optical wavelengths, the large dash line to 2-10 keV X-rays,
the small dot-dash line to .1-30 GeV \protect$\gamma$-rays while the large dot-dash line
to  > 500 GeV $\gamma$-rays.}
\end{figure}

\begin{figure}[t]
\epsfxsize=10.2 cm
\epsffile{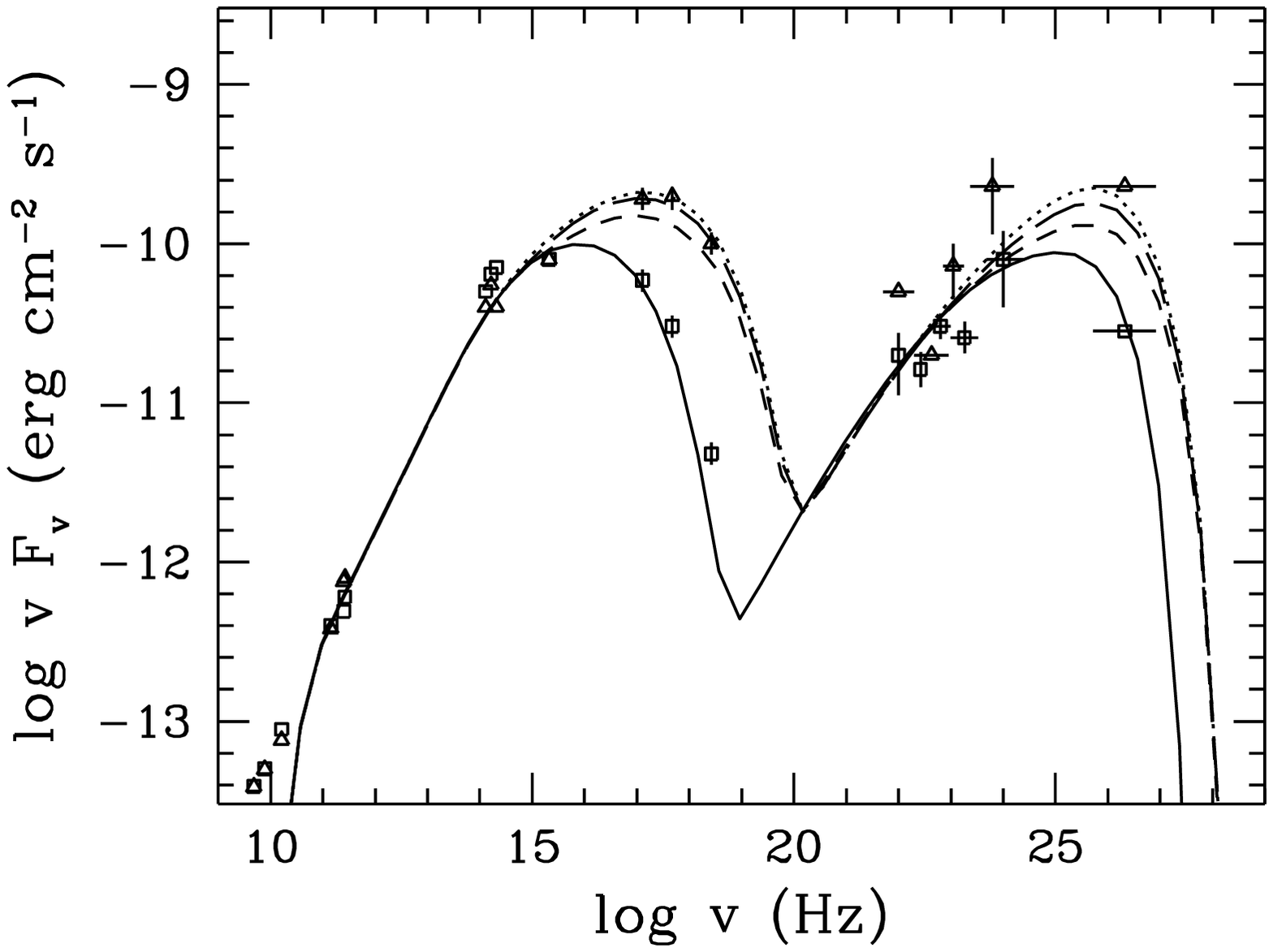}
\vspace{-2.5 cm}
\caption{\protect\label{gmaxplot}
Evolution of the multiwavelength spectrum of Mkn~421 in the case where the 
maximum energy of the electrons \protect$\gammamax$ changes impulsively by a 
factor of 5.
The solid line corresponds to the quiescent spectrum of Fig.~1. The short
and long dashed
lines show the spectrum at 1 and 2 \protect$\tvar$ after the change in the electron
injection (corresponding to 1.2 and 2.4 days respectively).
The dotted line shows the new steady state.}
\end{figure}
\begin{figure}
\epsfxsize=10. cm
\epsffile{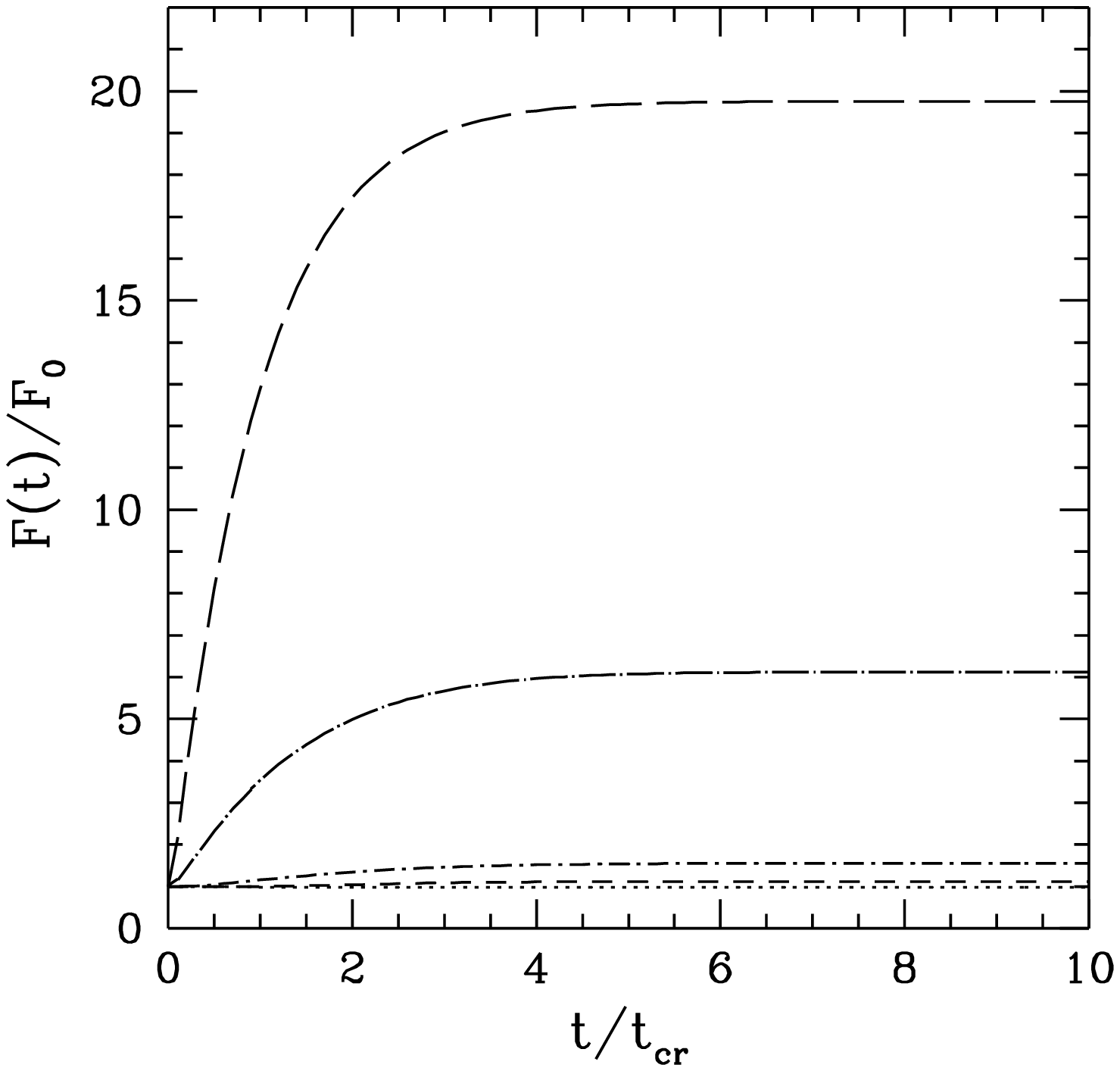}
\caption{\protect\label{lightgmax}
Plot of the flux at various frequencies (normalised to its quiescent value)
for the flare that corresponds to a change in \protect$\gammamax$ by a factor of 5.
The dotted line corresponds to a  wavelength of 0.4 cm,
the small dash line to optical wavelengths, the large dash line to 2-10 keV X-rays,
the small dot-dash line to .1-30 GeV \protect$\gamma$-rays while the large dot-dash line
to  > 500 GeV $\gamma$-rays.}
\end{figure}

\begin{figure}[t]
\epsfxsize=10. cm
\epsffile{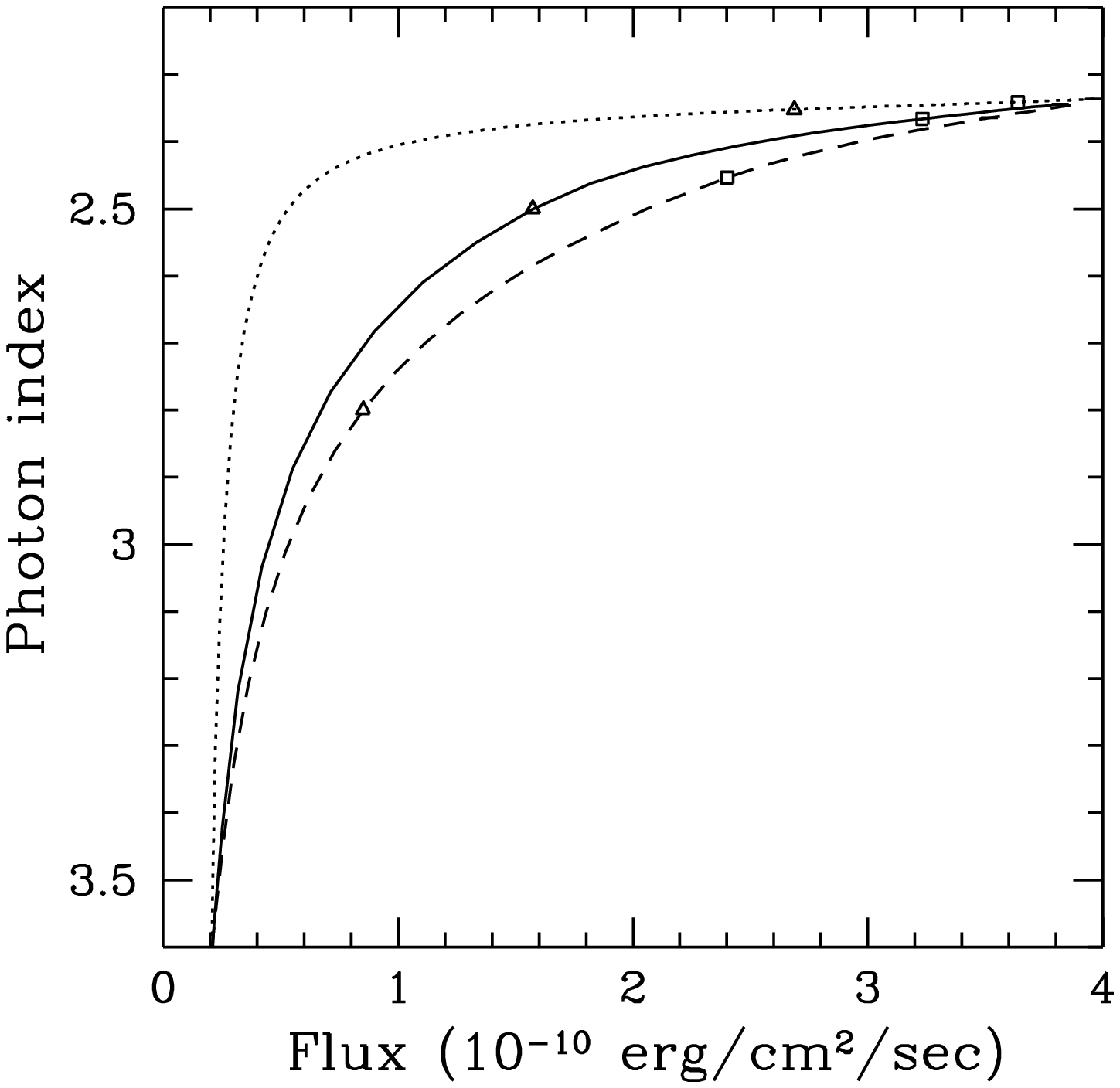}
\vspace{-1 cm}
\caption{\protect\label{slopeint}
Plot of the 2-10 keV photon index as a function of the flux at the same 
energy band that corresponds to a change in \protect$\gammamax$ by a factor of 5.
The dotted line corresponds to an impulsive change in \protect$\gammamax$,
while the full and dashed lines correspond to a more gradual change
with the change completed in the first case in one 
\protect$\tvar$ and in $2\protect\tvar$ in the second.
The triangles and squares indicate the values at 1 and 
$2\protect\tvar$ respectively.}  
\end{figure}
\begin{figure}
\epsfxsize=10. cm
\epsffile{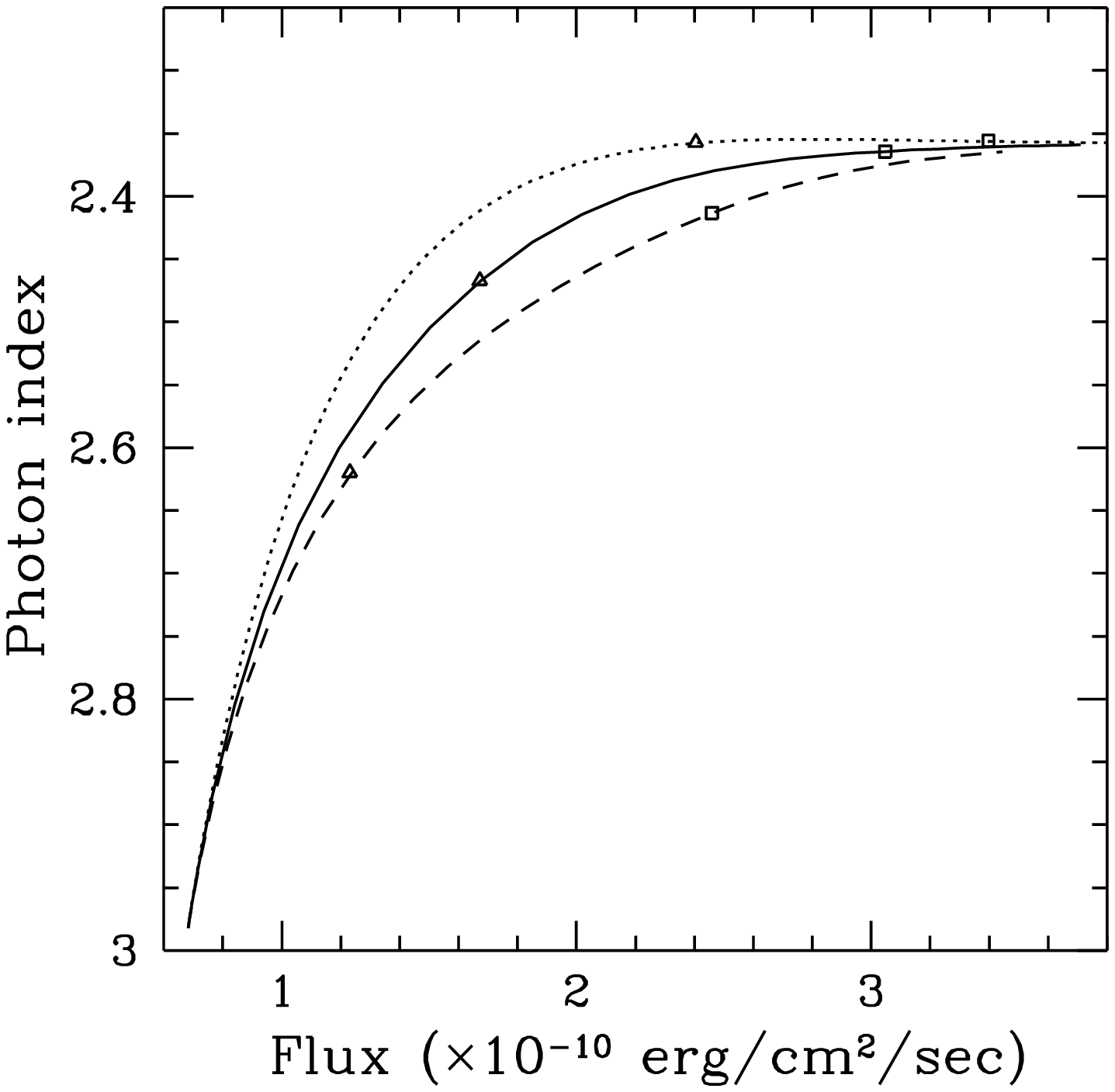}
\vspace{-1 cm}
\caption{\protect\label{slopetev}
Plot of the >500 GeV photon index as a function of the flux at the same 
energy band that corresponds to a change in \protect$\gammamax$ by a factor of 5.
The dotted line corresponds to an impulsive change in $\protect\gammamax$,
while the full and dashed lines correspond to a more gradual change
with the change completed in the first case in one 
$\protect\tvar$ and in $2\protect\tvar$ in the second.
The triangles and squares indicate the values at 1 and 
$2\protect\tvar$ respectively.}  
\end{figure}
\subsection{The flaring state}

In the present model, the simplest way to explain the flaring activity 
of Mkn~421 reported by Macomb et al. (\cite{macomb95}) 
is to change the electron 
parameters $\lowq$ and/or $\gammamax$ or to change the magnetic field. 
The first two can be understood as sudden variations in the acceleration 
mechanism while the third corresponds to the blob entering a region of 
higher field strength.

\subsubsection{Flares due to changes in $\lowq$}

Fig.~\ref{qinjplot} shows how a flare evolves in the case in which
the amplitude of the electron injection spectrum
$\lowq$ is suddenly increased by a factor of 3 above the quiescent 
value found in the previous section and then left constant.
The short and long dashed lines correspond to  snapshots of the
spectrum after 1 and $2\tvar$, respectively, measured from 
the time at which $\lowq$ changed ($\tvar\simeq 1.2$ days). 
The dotted line corresponds to the new steady state, which is
achieved many days later. It is evident that a change in $\lowq$
produces prompt flaring activity in the UV/X-ray and TeV $\gamma-$ray 
bands.
This can be attributed to the fact that high energy electrons have shorter
cooling times than low energy ones. Therefore the newly injected
high amplitude 
electron distribution cools first at the high frequency end, 
producing
synchrotron X-rays and inverse Compton  TeV gamma-rays. 
The lower frequency photons (optical synchrotron and GeV gamma-rays)
lag behind somewhat. This can be better  seen in 
Fig.~\ref{lightqinj}, which shows the lightcurves of various frequency
bands as a function of time (the travel crossing time across the emitting
region has not been taken into account).
The X-rays have the fastest 
response while the optical shows a slower response. 
Similarly TeV radiation has a faster response than GeV radiation. 
Note also that the amount by which the synchro-Compton component 
increases is larger than that by which the 
synchrotron component rises, since the latter varies approximately 
as proportional to $\lowq^2$ whereas
the former is roughly proportional to $\lowq$. 
\footnote{In the present
model self-consistent
electron cooling is included by both synchrotron and inverse Compton losses.
This breaks the simple linear dependence of synchrotron flux on $\lowq$.}
However, this holds only if the
change in $\lowq$ lasts long enough for a new quasi steady state to be 
established. If, for example, the electron injection is turned off after
one crossing time, this effect will not be observed. 
Finally,  it can be seen from 
Fig.~\ref{qinjplot},  that although the predicted flare 
matches the observed increase in amplitude as observed by Whipple,
it does not reproduce the hard X-ray spectral shape.
In addition, the model underestimates the high X-ray
flux by at least one order of magnitude.

\begin{figure}[t]
\epsfxsize=10.2 cm
\epsffile{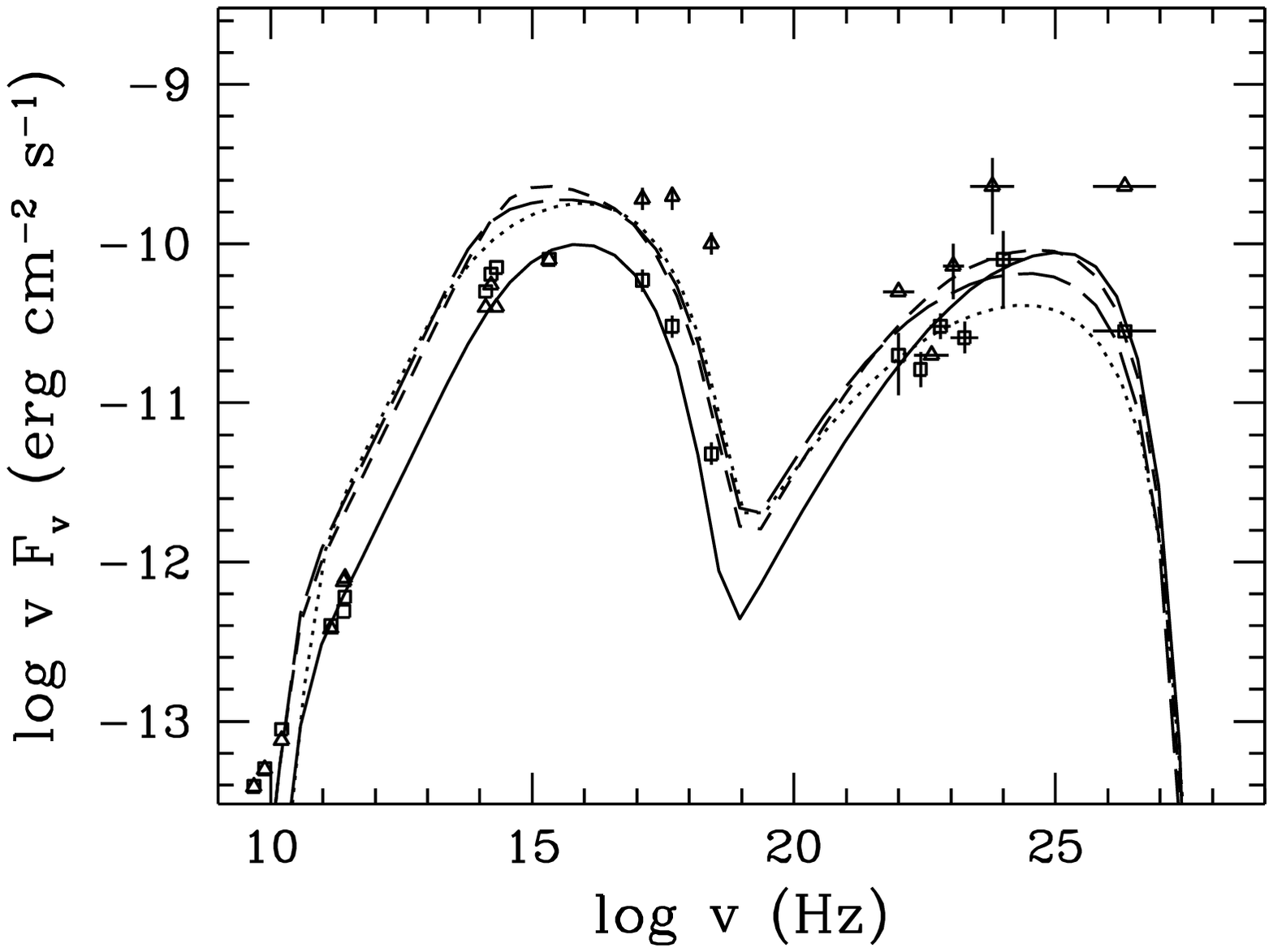}
\vspace{-2.5 cm}
\caption{\protect\label{bfplot}
Evolution of the multiwavelength spectrum of Mkn~421 in the case where the 
magnetic field strength changes impulsively by a factor of 3 and then
left constant.
The solid line corresponds to the quiescent spectrum of Fig.~1. The short
and long dashed
lines show the spectrum at 1 and 2 \protect$\tvar$ 
after the change in the electron
injection (corresponding to 1.2 and 2.4 days respectively).
The dotted line shows the new steady state.}
\end{figure}
\begin{figure}
\epsfxsize=10. cm
\epsffile{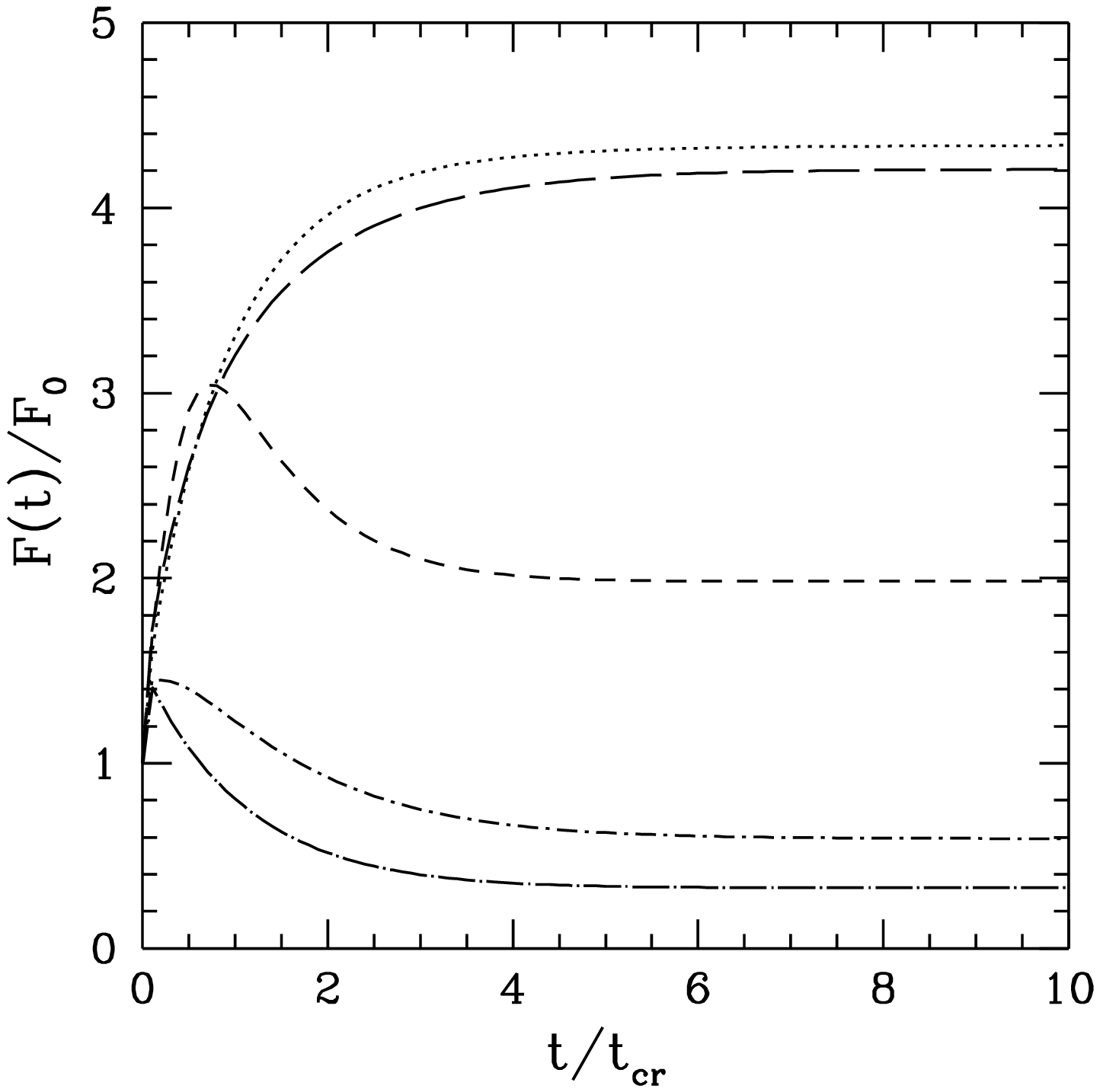}
\caption{\protect\label{lightbf}
Plot of the flux at various frequencies (normalised to its quiescent value)
for the flare that corresponds to a change in magnetic field strength by a factor of 3.
The dotted line corresponds to a  wavelength of 0.4 cm,
the small dash line to optical wavelengths, the large dash line to 2-10 keV X-rays,
the small dot-dash line to .1-30 GeV \protect$\gamma$-rays while the large dot-dash line
to  > 500 GeV $\gamma$-rays.}
\end{figure}

\subsubsection{Flares due to changes in $\gammamax$}

The evolution of a flare which corresponds to an impulsive increase
of $\gammamax$ by a factor of 5 
is shown in Fig.~\ref{gmaxplot}. Once again 
the short and long dashed lines correspond to  snapshots of the
spectrum after 1 and 2~$\tcross$ respectively as measured from the 
time of the sudden change
in $\gammamax$. The final steady state is shown as a dotted line
and it is reached after roughly 3$\tcross$. In contrast to the 
previous case, the predicted 
flaring activity due to an increase in $\gammamax$
gives a good fit both in X-rays and TeV $\gamma-$rays. 
Figure~\ref{lightgmax} displays the light curves predicted in  
frequencies from infra-red to hard gamma-rays. 
The evolution of the flare is very different from the previous case 
as here the source exhibits 
an outburst only in X-rays and TeV $\gamma-$rays while 
the other frequencies remain practically unaffected. 
This behaviour was also found by Marscher \& Travis 
(\cite{marschertravis96}) and it was suggested by 
Macomb et al.~(\cite{macomb95})
as a possible interpretation of their data.
Note that the variability in the GeV range is of the same
order as the size of the error bars.
Furthermore we find that, contrary to the previous case, the X-ray flare
is of higher amplitude than the TeV one.

  From Fig.~\ref{lightgmax} it is also evident that in this case
the outburst is strongest in the X-ray regime.
In fact, there are marked changes in the spectral index of the X-rays 
during a flare. This is displayed in Fig.~\ref{slopeint}, which plots 
the $2$--$10\,$keV spectral  index as a function of flux in this energy 
range. Three different flares are plotted, which differ in the time 
scale over which $\gammamax$ is increased: first of all impulsively
(dotted line), then on a time scale of $\tvar $ (solid line)
and then $2\tvar$ (dashed line). In each case
a spectral hardening in X-rays with increasing flux is
predicted, in qualitative agreement
with recent ASCA observations (Takahashi et al.~\cite{takahashietal96}). 
A similar effect can be observed in the hard gamma-ray 
($>500\,$GeV) range. This is 
shown in Fig.~\ref{slopetev}. 

\subsubsection{Flares due to changes in $B$}

As a final example, we present a flare caused by a sudden increase
in the strength of the magnetic field. 
Fig.~\ref{bfplot} depicts the spectral 
evolution that corresponds to an impulsive increase
of the magnetic field strength by a factor of 3 
and then left there. In this case, 
the whole synchrotron spectrum is shifted to the right by the same factor and
is also boosted in intensity, as might be expected. 
The synchro-Compton component, on the
other hand, is reduced compared to its original 
steady state value. The reason is that the ratio
$\elcomp/\db$ is reduced when the field increases, 
so that the total luminosity (which is held constant)
is redistributed towards the synchrotron component. This behaviour can also 
be seen in Fig.~\ref{lightbf} where one can observe a fast increase in the 
flux of low frequency bands along with a decrease of the flux in the 
$\gamma$-ray bands. This result contrasts strongly with
similar investigations by Bloom \& Marscher~(\cite{bloommarscher96}).
There is, however, no discrepancy, since these authors consider
a change in the magnetic field whilst  holding the electron 
{\em distribution} constant. In our case, we keep the electron
injection (and hence the total luminosity) constant. 

\section{Discussion}

In the present paper we obtained fits to the X-ray/TeV flares
observed during the 1994 multiwavelength campaign of Mkn~421
in the context of the homogeneous synchro-Compton model.
This model assumes that the most important photon targets for 
inverse Compton scattering by relativistic electrons are the
synchrotron photons they themselves produce. In the case of BL~Lac 
objects  such as Mkn~421, this assumption may be justified, since there
is no evidence of a more important photon component. However, this may 
not be the case in other sources, where photons from an accretion disk 
(Dermer, Schlickeiser \& Mastichiadis~\cite{dermeretal92}) or photons 
scattered from broad line clouds (Sikora, Begelman \& Rees~\cite{sikoraetal94})
may dominate. 
Nevertheless, a combination of synchrotron
radiation and inverse Compton scattering on external photons
(Dermer, Sturner \& Schlickeiser 
~\cite{dermeretal96}) 
may be useful in modelling results such as those of the multi-wavelength
campaign on 3C273 (Lichti et al.~\cite{lichtietal95}).
Whether a homogeneous synchro-Compton model such as presented here
can explain these or similar observations (e.g., those of 3C279: Maraschi et al.~\cite{maraschietal94}, 
Hartman et al.~\cite{hartmanetal96}) is currently under
investigation.

We obtain the full time dependent behaviour of flares by fitting the 
`quiescent' spectrum of the source and then varying one of the free 
parameters. Three simple ways of producing flares were investigated: 
(i) by changing the amplitude of the electron injection spectrum, (ii) 
by changing the maximum energy of the electron injection spectrum and 
(iii) by changing the magnetic field strength. We found a good fit to 
the observations using a flare of type (ii). This produces changes only 
in the X-ray and TeV bands, leaving all the other bands essentially 
unaffected. It also reproduces qualitatively the observed hardening of 
the X-rays with increasing intensity (Takahashi et al.~\cite{takahashietal96}). 
We also found that X-rays are first to react to any change
in the electron injection. This is particularly pronounced for flares of
type (i) and (ii) (see Figs.~\ref{lightqinj} and \ref{lightgmax}).
A change in the acceleration parameters is tracked more closely by
X-ray photons than by photons in other wave-bands, since 
the X-ray producing electrons
have the highest energy and, therefore, the 
the fastest cooling timescale.
	
The parameters we find for the fits are similar to those found 
by other authors (see, for example, Marscher \& 
Travis~(\cite{marschertravis96}),
Sambruna, Maraschi \& Urry~(\cite {sambrunaetal96})).
The fast time variability of Mkn~421 ($\simeq$1 day) implies
a Doppler factor of $\simeq$15. However, 
a faster variation can easily be accommodated in our model, since 
new fits with a larger Doppler factor can be obtained simply scaling 
the parameters as indicated in Sect.~3. One potentially independent 
constraint on  the model is provided by the effects of synchrotron 
self-absorption. At the lower radio frequencies, the absorption 
turnover can be seen in Figs.~\ref{quiet}, \ref{qinjplot}, \ref{gmaxplot} 
and \ref{bfplot}. For higher Doppler factors, 
this effect should disappear, since the required luminosity is then 
provided by a lower value of $\elcomp$. As a result, the electron 
column density of the source is reduced. However, the homogeneous SSC 
model presented here is so compact that the effects of induced  Compton 
scattering can also be expected to manifest themselves in the radio  
range (Coppi, Blandford \& Rees~\cite{coppietal93}, Sincell \& 
Krolik~\cite{sincellkrolik94}). 
A simple estimate of the importance of this effect can be obtained 
by evaluating the parameter $\tau_{\rm ind}=
N_{\rm e}\sigmaT R T_{\rm br}/(\melec c^2)$,
where $T_{\rm br}$  is the 
brightness temperature of the radiation in 
energy units and $N_{\rm e}$ is the total electron density in the source.  
Even if we assume that no thermal electrons are present, this parameter exceeds
unity for frequencies less than roughly $700\,$MHz, given the 
parameters of our quiescent model. Thermal electrons, however, 
accumulate within the source, 
although we have not calculated their density explicitly. consequently, 
a modification of the simple synchrotron spectrum at gigaherz frequencies 
may be possible.

Our calculations predict the time-dependent flaring activity to be 
expected given certain idealised fluctuations in the injection spectrum 
of high energy electrons into a relativistically moving blob. 
Although we do not address a specific acceleration  mechanism, it is possible to 
interpret the overall picture as one in which a shock front rapidly 
accelerates electrons and leaves a trail of them (i.e., injects them) 
in the plasma streaming away from the shock on its downstream side. In 
our model, we assume the typical dimension of the radiating part of the 
downstream plasma in its rest frame is $R$. The photon escape time is 
$\tcross$, which we find to be roughly one third of the time taken for 
electrons to cross the emitting region. Thus, assuming electrons are 
carried along by the plasma, our picture would indicate rapid ($\sim 
c/3$) movement of the downstream plasma away from the shock front. In a 
more  realistic model, the spectrum of injected  electrons should also  
be calculated in a time-dependent manner.  In fact, the value of 
$\gammamax$ itself should probably be determined by a balance between 
acceleration and losses in a self-consistent model. This is possible in 
a hadronic model (e.g., MK95), however, self-saturation  of accelerated 
protons requires a very high photon density, which would render the 
source opaque to TeV gamma-rays.  On the other hand, it is possible to 
imagine a model in which protons saturate at extremely high energy 
(e.g., Mannheim~\cite{mannheim93}), and inject electrons which 
produce the entire emitted spectrum by the synchrotron mechanism. The 
time-dependent properties of flares from such a model would, however, 
differ strongly from those found above. A detailed investigation of 
electron acceleration in the presence of losses (Heavens \& 
Meisenheimer~\cite{heavensmeisenheimer87}) has so far been performed 
only in the steady state, and without considering  either bulk 
relativistic motion or inverse Compton scattering. 

Another simplification we have introduced comes from the fact that we
have used a spherical homogeneous model (for a discussion of the
problems inherent with this model see, for example, Bloom \& Marscher 
\cite {bloommarscher96}). However this allows us to understand better
the significance of the various physical quantities we are using and,
at least for the case of Mkn~421, it proved adequate to fit the
multiwavelength spectrum. On the other hand, the inhomogeneous
models might be superior to
the homogeneous ones in the sense that they give better overall fits
to the AGN spectra, however they introduce a number of extra parameters
making a simple understanding of the results difficult.

\acknowledgements
We would like to thank an anonymous referee who helped us
clarify many of the issues presented here.
AM thanks the Deutsche Forschungsgemeinschaft for support under
Sonderforschungsbereich 328.

\end{document}